# High brightness single mode source of correlated photon pairs using a photonic crystal fiber

J. Fulconis[1], O. Alibart[1], W. J. Wadsworth[2], P. St.J. Russell[2] and J. G. Rarity[1]

[1]*Centre for Communications Research, Department of Electrical and Electronic Engineering, University of Bristol,, Merchant Venturers Building, Woodland Road, Bristol, BS8 1UB, United Kingdom.*
*Jeremie.Fulconis@bristol.ac.uk*
[2]*Photonics & Photonic Materials Group, Department of Physics, University of Bath, Claverton Down, Bath, BA2 7AY, UK*

**Abstract:** We demonstrate a picosecond source of correlated photon pairs using a micro-structured fibre with zero dispersion around 715 nm wavelength. The fibre is pumped in the normal dispersion regime at ~708 nm and phase matching is satisfied for widely spaced parametric wavelengths. Here we generate up to $10^7$ photon pairs per second in the fibre at wavelengths of 587 nm and 897 nm. On collecting the light in single-mode-fibre-coupled Silicon avalanche diode photon counting detectors we detect $\sim 3.2 \cdot 10^5$ coincidences per second at pump power 0.5 mW.

©2005 Optical Society of America

## 1. Introduction

High brightness sources of correlated and entangled photon pairs are required for various multiphoton and linear optical logic applications [1-3]. They also find useful application to quantum cryptography [4] and quantum communications [5]. The preferred sources for such experiments until recently have been three wave mixing in $\chi^{(2)}$ non-linear birefringent crystals [6]. These sources are inherently wide band, low brightness (per nanometer, per single mode) sources and suffer from mode matching problems when coupling to optical fibre [7]. More recently periodically poled fibres [8] and periodically poled waveguides of lithium niobate have been shown to be useful pair photon sources [9]. In poled fibres the low non-linearity limits the brightness while in planar waveguides the non-circular mode limits the coupling efficiency into optical fibres.

It is well known that parametric gain can arise from the $\chi^{(3)}$ non-linearity in optical fibres [10,11] and phase matching can be achieved by using the modulation instability when pumping fibres in their anomalous dispersion regime. Various pair photon generation experiments have been performed in this regime [12-16]. The photon pairs are generated close to the pump wavelength and are always accompanied by a significant Raman background and careful filtering is required.

We have recently shown that phase matching can be obtained for widely spaced wavelengths by pumping photonic crystal fibre (PCF) close to the zero dispersion wavelength in the *normal* dispersion regime [17]. This allowed us to demonstrate a CW source of photon pairs at widely spaced wavelengths of 834 nm signal and 1404 nm idler by pumping at 1047nm wavelength [18, 19]. However the idler signal lies on the shoulder of the fifth order Raman peak which is still a significant source of background light. Here we choose fibre with the zero dispersion point in the near infra-red (715nm) and pump with a *picosecond pulsed* laser, red detuned a few nanometers into the normal dispersion regime. This generates photon pairs visible to efficient silicon-based photon counting detectors and since the number of created photon pairs is proportional to the square of the peak intensity while the Raman scattering grows roughly linearly, the use of a picosecond laser in our experiment allows us to improve the brightness but also to reduce the Raman source of background noise to negligible levels.

We are seeking to develop a source which may be applicable for future quantum interference experiments involving three or more photons created as two or more pairs. Interference effects between separate pair-photons can be studied by overlapping photons with a time uncertainty shorter than their inverse bandwidth or coherence length [20]. This restricts us to sources pumped by ultra-short laser pulses where the bandwidth is of order nanometers and also requires a high efficiency of collection. We show in the following that our source is naturally narrow band (3-6 nm) and is efficiently coupled into single mode optical fibers, thus being ideal for quantum interference experiments.

## 2. Theory

Here the main nonlinear process that has to be taken into account is four-wave mixing (FWM) where phase matching and conservation of energy give the equations (1) and (2) [10]:

$$k_i + k_s - 2k_p + 2\gamma P_p = 0 \qquad (1)$$

and

$$\omega_i + \omega_s = 2\omega_p \qquad (2)$$

where $k_{i,s,p}$ are the wave-vectors (propagation constants) of the idler, signal and pump photons and $\omega_{i,s,p}$ their respective frequencies; $P_p$ is the peak pump power and $\gamma$ is the nonlinear coefficient of the fiber,

$$\gamma = \frac{2\pi n_2}{\lambda A_{eff}} \qquad (3)$$

where $n_2 = 2 \times 10^{-20}$ m²/W is the nonlinear refractive index of silica, $A_{eff}$ is the effective area of the fiber mode and $\lambda$ is the pump wavelength. These phase-matching conditions will yield the wavelengths for peak gain in a given fiber, which will depend on the chromatic dispersion of the fiber.

Figure 1 shows the microstructured fiber used in our experiment. The fiber has a core diameter of 2 μm and a zero dispersion wavelength $\lambda_0$ defined at 715 nm. Figure 2 shows the phase matching diagram plotting signal and idler wavelength against the pump. This is calculated from equations (1) and (2) using the dispersion curve for a simple strand of silica in air as a close approximation [21] to the fiber used in this experiment. A strand diameter of 2 μm was used which gives a zero dispersion wavelength at 715 nm as measured in the fibre.

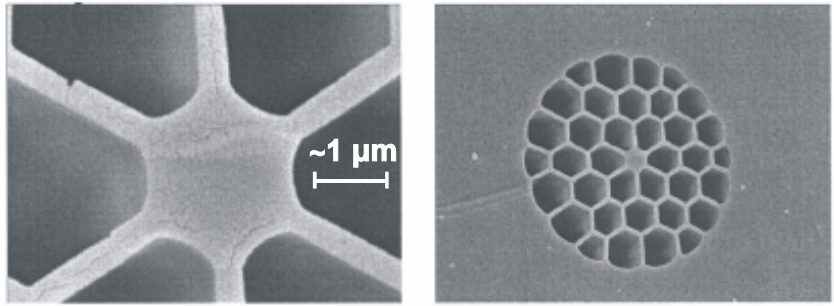

Fig. 1. Electron microscope image of the PCF used with core diameter $d \approx 2$ μm, $\lambda_0 = 715$ nm

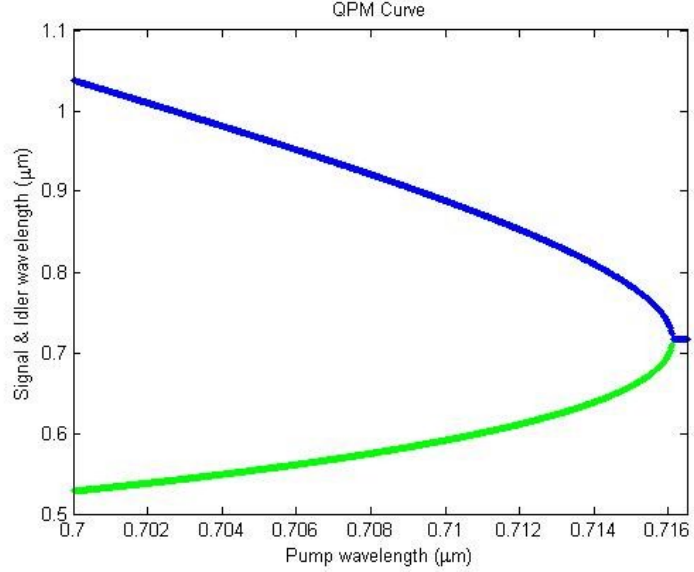

Fig. 2. Nonlinear phase-matching diagram for the process $2\omega_p \to \omega_s + \omega_i$. The curve does not change significantly in the range $P_p = $ 0-3 W.

Since the number of created photon pairs is proportional to the square of the peak intensity while the Raman scattering grows roughly linearly, the use of a picosecond laser in our experiment allows us to improve the brightness but also to reduce the Raman source of background noise to negligible levels. For a given average power from the pump laser, the pair photon signal will scale as the inverse of the mark to space ratio of the pulses, whereas the Raman signal will be unchanged.

In our experiments we pump in the normal dispersion regime, where the sidebands generated are widely spaced at equal frequency intervals from the pump. The natural linewidth determined from the phase matching relations (when pumped by a monochromatic source) is below 0.2 nm. Hence we expect the bandwidth of parametric light generated by a picosecond pumped source to be mainly determined by the slope of phase matching curves multiplied by the pump bandwidth. Here we use a pulse full width half maximum (FWHM) bandwidth of 0.3 nm (corresponding to a ~4 ps pulse) centered at 708.4 nm, which corresponds to a calculated FWHM bandwidth of 3.2 nm at 587 nm for the signal and 4.5 nm at 897 nm for the idler.

### 3. Experiment

In order to estimate the brightness of our source we used the coincidence setup depicted in Fig. 3 where a mode-locked picosecond Ti:Sapphire pump laser (Spectra Physics - Tsunami) set at 708.4 nm, emitting 4 ps pulses with a repetition rate of 80 Mhz is sent, through an optical isolator, onto a prism P to remove in-band light from the pump laser spontaneous emission. A pin hole is then used to improve the pump mode and eventually several attenuators bring the power down so that up to 540 µW average power is launched into the fiber. Since the PCF is birefringent and supports two modes, a half wave-plate (HWP) is used to align the pump polarization along one axis thus preventing polarization scrambling and creating pairs with the same polarization as the pump beam. Compared to bulk crystal where the nonlinear interaction occurs actually at the pump beam focus, here the interaction will extend over the full length of the fiber and is thus more efficient. We then used for our experiment 2 m of PCF allowing us to expect a high conversion efficiency. The output of the fiber is collimated using an aspheric lens, followed by a removable mirror M, allowing us to monitor the photon pair spectra in a monochromator or to launch them into a coincidence test bench. In this latter part of the setup, a dichroic mirror centered at 700 nm is used to spread the incoming beam into two arms, one corresponding to the *signal* channel and the other to the *idler*, where band-pass filters F1 and F2 centered at 570 nm and 880 nm respectively (width ~ 40 nm, T>80%) are used to remove in-line pump and background light. Each photon of the pair is then launched into single mode fibers that are connected to two Silicon avalanche photodiodes (APD). The detected photons are counted in a dual-channel counter and the coincidences between the two APDs are analyzed using a time interval analysis system (TIA).

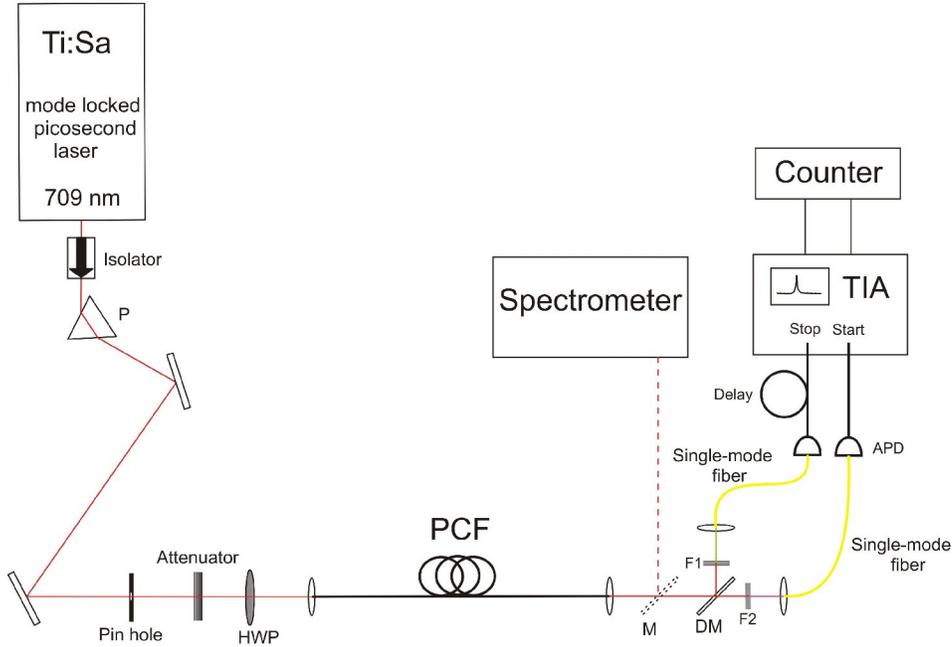

Fig. 3. Optical layout. Laser, 708 nm Ti:Sa laser; P, prism; HWP, halfwave plate; PCF, 2 m of photonic crystal fiber; M, protected silver mirror (R>95%); DM, dichroic mirror (centered@700nm, T>85%, R>90%); F1, 570 nm band-pass filter, bandwidth 40 nm, T=80%; F2, 880 nm band-pass filter, bandwidth 40 nm, T=80%; APD, Silicon single photon detector.

## 4. Results

Pumping with 708.4 nm light and aligning the polarisation on one of the axes of the fiber, we see the narrowband pair-photon spectra illustrated in fig.4 and fig. 5. The short wavelength sits at 587 nm and the corresponding idler at 897 nm. Note that these wavelengths are easily tunable using fiber parameters or pump wavelength thus allowing us to set the signal and idler photon far from the degeneracy point ($\lambda_0$=715 nm) where the noise from Raman background is negligible. The pair emission is narrow band and we measure 2.7 nm and 5.5 nm FWHM bandwidths for the signal and idler respectively. These correspond well with the simple theory presented in section 2. The spectra were taken at a pump power of 700 μW and flat background comes entirely from the electronic bias errors in the CCD. In the IR signal we see a very slight increase in background towards shorter wavelengths which we ascribe mainly to residual Raman scattering.

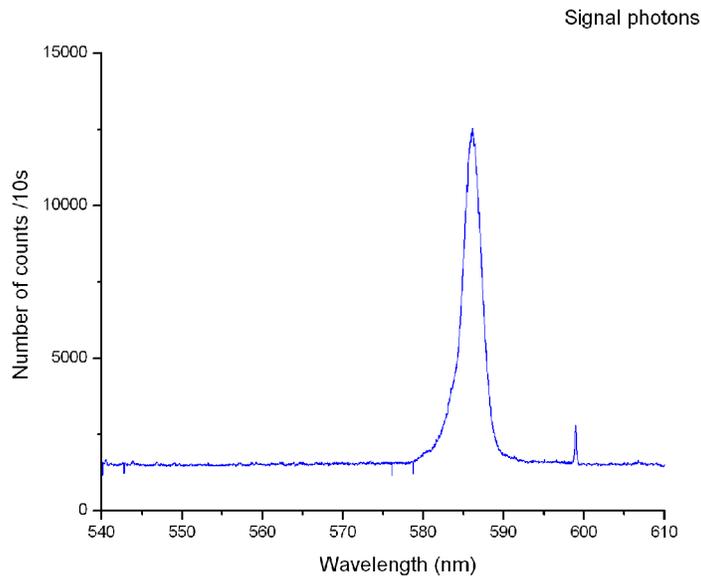

Fig.4: Fluorescence spectrum of the signal photons. The measurement was integrated over 10 seconds. The number of counts is proportional to the number of photons detected by the cooled camera. The photon spectrum is centered at 587 nm and features a FWHM bandwidth of 2.7 nm.

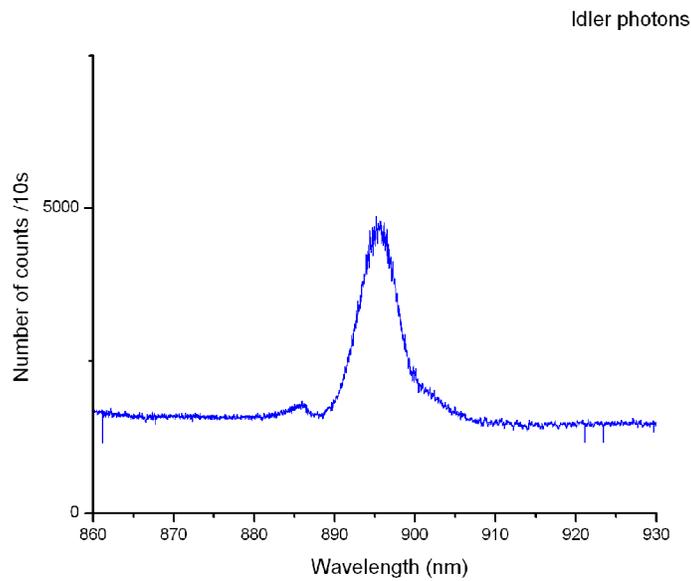

Fig.5: Fluorescence spectrum of the idler photons. The measurement was integrated over 10 seconds. The photon spectrum is centered at 897 nm and features a FWHM bandwidth 5.5 nm.

Looking now at the coincidence test bench in order to determine the brightness of our source, we measured the number of single counts in both signal and idler channels, while we recorded the number of coincidences. This experimental protocol amounts to recording the percentage of detections in the "start channel" which have been stopped by detection in the "stop channel" in the following time interval and effectively provides a direct estimate of the 'lumped' efficiency of the stop channel.

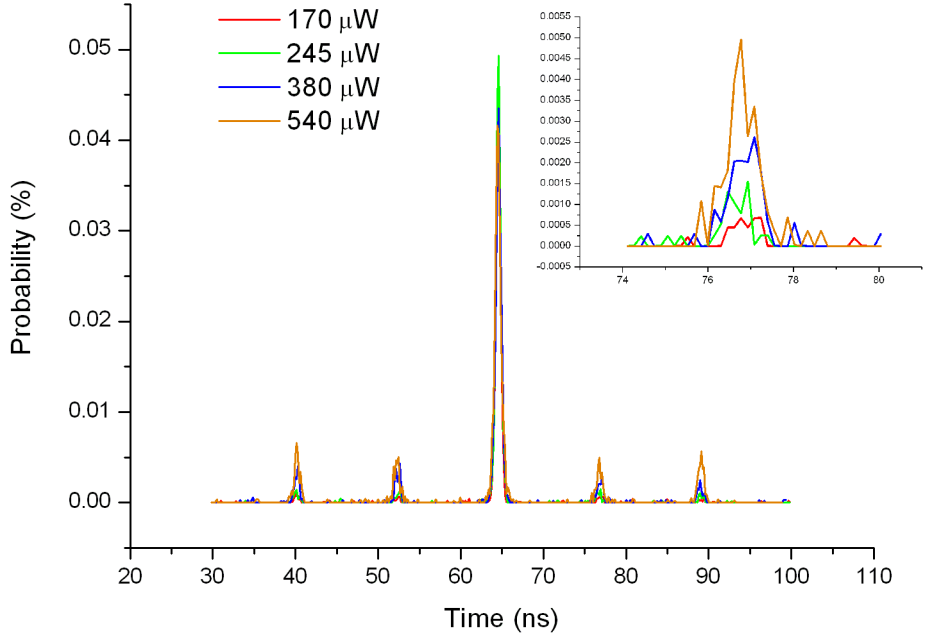

Fig.6: Time interval histogram showing the coincident photon detection peak and also a zoom on one of the accidental coincidence peak for different pump powers. Here the time between two peaks reflects the pump laser repetition rate. However the width of the peaks is limited by the response time of the detectors which is typically hundreds of picoseconds (rather than the actual duration of the pump pulses). To step through the pump powers press Ctrl + click on the graph.

The central peak corresponds to signal and idler photons belonging to the same pulse, whereas the small satellite peaks stand for uncorrelated events i.e. signal and idler coming from different pulses. It is interesting to note the satellite peaks grow with pump power, whereas the central peak remains constant. In this context, the central peak is the proof of correlated photon pair creation, while the satellite ones are linked to the creation of multi-photon pairs and Raman noise. However we also have to take into account this "background" rate in the central peak, thus we will introduce $C_{raw}$ as the raw coincidence rate in the central peak and we will calculate the accidental coincidence rate thanks to the satellites peaks $C_b$.

We can use the singles counting rates and the coincidence rates to estimate the actual rate of pairs created inside the PCF, using the following equations [22]:

$$N_s = \eta_s \eta_{opt} r + B_s$$
$$N_i = \eta_i \eta'_{opt} r + B_i \quad (4)$$
$$C_{raw} = \eta_s \eta_i \eta_{opt} \eta'_{opt} r + C_b$$

where $N_s$, $N_i$ are respectively the counting rates in the signal and idler APDs, while $B_s$, $B_i$ are background rates including Raman photons and multi-photon pairs. $\eta_s$ and $\eta_i$ are the APD quantum efficiencies at 587 nm and 897 nm, the net optical transmission and launch efficiencies into single mode fibre of each arm are $\eta_{opt}$ and $\eta'_{opt}$. If we neglect the background we can directly estimate the measured lumped efficiencies $\eta_M^{lump}$ of each arm and the pair photon production rate $r$

$$\eta_{sM}^{lump} = \frac{C_{raw} - C_b}{N_i}$$
$$\eta_{iM}^{lump} = \frac{C_{raw} - C_b}{N_s} \quad (5)$$
$$r = \frac{C_{raw} - C_b}{\eta_{iM}^{lump} \eta_{sM}^{lump}}$$

We can also estimate the efficiencies directly. From the detector data sheet [23] we can estimate $\eta_s \sim 0.6$, and $\eta_i \sim 0.33$ at 587 nm and 897 nm respectively. The launch efficiencies into single mode fibre of each arm are estimated to be 60-65% and when combined with the measured transmission of lenses and filters we estimate $\eta_{opt} \sim 0.35\text{-}0.38$, $\eta'_{opt} \sim 0.33\text{-}0.36$ (including filter, dichroic mirror and lenses losses for signal and idler channels respectively). This then gives us an estimate of $(\eta_s \eta_{opt})_P \sim 0.211\text{-}0.229$ and $(\eta_i \eta'_{opt})_P \sim 0.11\text{-}0.119$ where the subscript $P$ denotes predicted. We can then estimate the background contributions by combining (4) and (5) to give

$$\frac{\eta_{sM}^{lump}}{(\eta_s \eta_{opt})_P} = \left(1 - B_i/N_i\right)$$
$$\frac{\eta_{iM}^{lump}}{(\eta_i \eta'_{opt})_P} = \left(1 - B_s/N_s\right) \quad (6)$$

We then took several measurements for a range of different pump powers so that the ratio of Raman background to signal would change thus changing the ratios $B/N$ in equations (6) and compared the measured and predicted efficiencies. The results are shown in the following table 1.

Table 1. Summary of results for different pump powers

| Pump Power P | 170 μW | 245 μW | 380 μW | 540 μW |
|---|---|---|---|---|
| $N_S$ | $3.4\times10^5$ s$^{-1}$ | $6.8\times10^5$ s$^{-1}$ | $1.57\times10^6$ s$^{-1}$ | $2.89\times10^6$ s$^{-1}$ |
| $N_i$ | $1.9\times10^5$ s$^{-1}$ | $3.6\times10^5$ s$^{-1}$ | $8.2\times10^5$ s$^{-1}$ | $1.52\times10^6$ s$^{-1}$ |
| $C_{raw}$ | $3.9\times10^4$ s$^{-1}$ | $8.0\times10^4$ s$^{-1}$ | $1.8\times10^5$ s$^{-1}$ | $3.6\times10^5$ s$^{-1}$ |
| $C_b$ | $0.1\times10^4$ s$^{-1}$ | $0.2\times10^4$ s$^{-1}$ | $0.1\times10^5$ s$^{-1}$ | $0.4\times10^5$ s$^{-1}$ |
| $C=C_{raw}-C_b$ | $3.8\times10^4$ s$^{-1}$ | $7.8\times10^4$ s$^{-1}$ | $1.7\times10^5$ s$^{-1}$ | $3.2\times10^5$ s$^{-1}$ |
| Signal lumped efficiency $\eta_{sM}^{lump}$ eq. (5) | 20.0 % | 21.7 % | 20.7 % | 21.1% |
| Predicted Signal efficiency $(\eta_s\eta_{opt})_P$ | 21.1-22.9% | 21.1-22.9% | 21.1-22.9% | 21.1-22.9% |
| $B_i/N_i$ from eq. (6) | 0.052-0.126 | 0-0.052 | 0.019-0.096 | 0-0.08 |
| Idler lumped efficiency $\eta_{iM}^{lump}$ eq. (5) | 11.2 % | 11.5 % | 10.8 % | 11.1 % |
| Predicted Idler efficiency $(\eta_i\eta'_{opt})_P$ | 11.0-11.9% | 11.0-11.9% | 11.0-11.9% | 11.0-11.9% |
| $B_s/N_s$ from eq. (6) | 0-0.059 | 0-0.034 | 0.018-0.092 | 0-0.067 |
| Photon pair production rate in the fibre $r$ (eq. (5)) | $1.7\times10^6$ s$^{-1}$ | $3.1\times10^6$ s$^{-1}$ | $7.6\times10^6$ s$^{-1}$ | $1.4\times10^7$ s$^{-1}$ |
| Average number of pairs per pulse | 0.021 | 0.039 | 0.095 | 0.18 |

From table 1 we see that we have measured up to $3.2\times10^5$ net coincidences per second. We also see that the lower estimated idler efficiency is below the measured value suggesting that the fibre launch efficiency is higher than our lower estimate at ~62%. Also this confirms that the background in the signal count rate (green) is very low (<5% of the total rate). We expect this as there is no Raman signal in the green. In the infra-red we estimate a background between 3% and 10% of the count rate for all pump powers. This is counter to our expectation of a reduction of background at higher pump powers. However this may simply be due to our rather crude analysis. We have taken no account of dead-time corrections in the detectors and time interval analyser and only a simple account of multi-photon effects. We are presently trying to make an independent measure of the background and expect to improve on these results, which confirm that background in the IR is smaller than 10% of signal for all powers measured.

## 5. Discussion

We know that the background coincidence rate at all count rates here is dominated by random overlap of more than one pair of photons. We thus do not present our data in terms of coincidence to background ratios but work only with the measured efficiencies of 20.7+/-0.3% in the signal (green) and 11.3+/-0.2% in the idler (IR). These efficiencies imply greater than 60% coupling efficiency into single mode fibre has been achieved. Since the background coincidence rate is linked to the multi-photon pair probability, it is indicative of the four photon coincidence rate we will get in future experiments and is a variable we want to maximise.

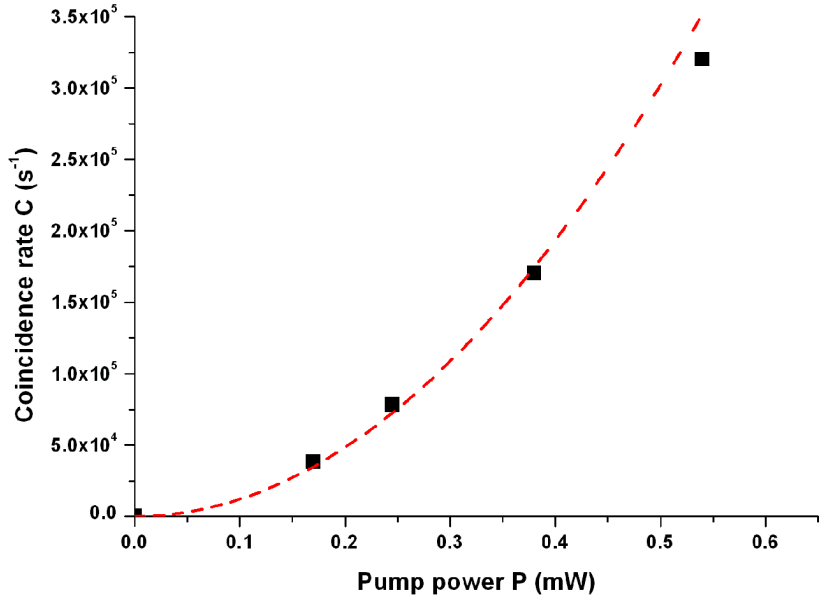

Fig. 7: Net coincidence rate as function of the pump power. The fit is purely quadratic (no linear term) $C=AP^2$ with constant $A=1.21 \times 10^6$ /sec/mW$^2$. Discrepancies at high powers are due to saturation effects in both the detectors and coincidence measuring apparatus.

Also when we plot on figure 7 the rate as a function of pump power we clearly see a purely quadratic behaviour. We cannot thus present our rates in terms of coincidences per milliwatt per nanometre. However we have seen 320,000 net coincidences per second with pump powers of 0.54 mW (or ~$1.2 \times 10^6$ /sec/mW$^2$) with natural bandwidths of signal 2.7 nm and idler 5.5 nm. This is this brightest source of pair photons and heralded single photons that we know of [24]. In addition to a high brightness, this source shows negligible background noise in the visible and in the IR the Raman scattering photons contribute less than 10% at all pump powers used. Our source is limited in brightness by two principle factors. The first is saturation of the detectors and coincidence measuring equipment, which is causing significant distortion at the highest pump powers. The second is the optical losses and detector efficiencies.

A preliminary experiment using a multimode fiber in the signal channel showed efficiency in the signal arm going up to 33% due to high coupling efficiency (>95%) in multimode fiber. This result highlights the limited launch efficiency into single-mode fibers, which we estimate to be about 62%. Although this value is already better than any nonlinear waveguide [9] solution to our knowledge, we expect this efficiency to be improved to >80% using a pigtailed fiber [25] which will compare favourably with the best results from bulk nonlinear crystal [7].

For quantum information we require a narrower bandwidth so that the coherence length is equivalent to the pulse length [20]. A quantum interference experiment involving four-fold coincidence between photons coming from two separated sources would require a filter of order 0.2 nm bandwidth in the Green (0.4nm in the IR). Such filters will transmit only 40% of in-band light thus halving our effective efficiencies and collect only ~1/15 of the available spectrum. However counting rates would be significantly reduced thus allowing an increase in pump power. Using our source with a pump power of 2 mW, the expected rate of photon pairs detected within this bandwidth is $>8.10^4$ /s, which means a rate of four photon events $>80$ s$^{-1}$, two orders of magnitude higher than any previous experiment.

## 6. Conclusion

We have reported the measurement of picosecond-pulsed photon pairs generated by four-wave mixing in a single-mode optical fiber, pumped in the normal dispersion regime. The source is polarized, bright, narrowband, single-mode and tunable by varying laser wavelength or fiber parameters. The wide separation of the generated pair wavelengths means that most of the background can be avoided. All these advantages make this new source of photon pairs more appropriated compared to conventional ones for quantum information processing applications.

## Acknowledgements

WJW is a Royal Society University Research Fellow. The authors would like to thank D. (ripper) Wardle for early input. The work is partly funded by UK EPSRC QIP IRC and EU IST-2001-38864 RAMBOQ.